%% file: Fac24.tex
\documentclass[sigconf,screen]{acmart}
\AtBeginDocument{%
  }

\setcopyright{none}
\copyrightyear{2024}
\acmYear{2024}

\acmConference[RecSys ’24]{FAccTRec 2024: The 7th Workshop on Responsible Recommendation. In 18th ACM Conference on Recommender Systems}{October 14--18, 2024}{Bari, Italy}

\begin{document}

\title{ARTAI: An Evaluation Platform to Assess Societal Risk of Recommender Algorithms}



\author{Qin Ruan}
\authornote{Both authors contributed equally to this research.}
\email{qin.ruan@ucdconnect.ie}
\orcid{0000-0001-5822-9260}
\author{Jin Xu}
\authornotemark[1]
\email{jin.xu@ucd.ie}
\orcid{0000-0002-6644-8217}
\affiliation{
    \department{School of Computer Science and Insight SFI Research Centre for Data Analytics}
    \institution{University College Dublin}
    \city{Dublin}
    \country{Ireland}
}


\author{Ruihai Dong}
\email{ruihai.dong@ucd.ie}
\orcid{0000-0002-2509-1370}
\affiliation{
    \department{School of Computer Science and Insight SFI Research Centre for Data Analytics}
    \institution{University College Dublin}
    \city{Dublin}
    \country{Ireland}
}

\author{Arjumand Younus}
\email{arjumand.younus@ucd.ie}
\orcid{0000-0001-7748-2050}
\affiliation{
    \department{School of Information and Communication}
    \institution{University College Dublin}
    \city{Dublin}
    \country{Ireland}
}

\author{Tai Tan Mai}
\email{tai.tanmai@dcu.ie}
\orcid{0000-0001-6657-0872}
\affiliation{
    \department{School of Computing}
    \institution{Dublin City University}
    \city{Dublin}
    \country{Ireland}
}

\author{Barry O'Sullivan}
\email{b.osullivan@cs.ucc.ie}
\orcid{}
\affiliation{
    \department{Insight SFI Research Centre for Data Analytics and School of Computer Science \& IT}
    \institution{University College Cork}
    \city{Cork}
    \country{Ireland}
}

\author{Susan Leavy}
\email{susan.leavy@ucd.ie}
\orcid{0000-0002-3679-2279}
\affiliation{
    \department{Insight SFI Research Centre for Data Analytics}
    \institution{University College Dublin}
    \city{Dublin}
    \country{Ireland}
}
\renewcommand{\shortauthors}{Ruan et al.}

\begin{abstract}
Societal risk emanating from how recommender algorithms disseminate content online is now well documented. Emergent regulation aims to mitigate this risk through ethical audits and enabling new research on the social impact of algorithms. However, there is currently a need for tools and methods that enable such evaluation. This paper presents ARTAI, an evaluation environment that enables large-scale assessments of recommender algorithms to identify harmful patterns in how content is distributed online and enables the implementation of new regulatory requirements for increased transparency in recommender systems.
\end{abstract}





\maketitle

\section{Introduction}
Societal risks emanating from how content is disseminated  by recommender algorithms in online platforms to increase engagement are now well documented~\cite{carroll2022estimating,milano2020recommender,Baker2024}. The Digital Services Act (DSA)~ \cite{eu2020dsa} currently being implemented across the European Union (EU) address these issues and mandates improved algorithmic transparency through third-party auditing. It also recognises the crucial role of research, granting access rights for independent researchers to evaluate the social impacts of digital platforms~\cite{GPAI24}. Many countries are also introducing online safety legislation to ensure the protection of children using online platforms and mandating increased levels of transparency (e.g.\cite{IRE,OSA}). To enable researchers and auditors to conduct risk assessments of recommender algorithms there is a need for the development of evaluation tools~\cite{mokander2022conformity}.  ARTAI (Assessing Risk for Trustworthy AI) is an evaluation platform to address this requirement to assess the social impact of recommender algorithms.

Current approaches to risk evaluation involve analysis of code, user surveys, data scraping, sock-puppet audits and crowd-sourcing. Code audits are increasingly infeasible given the use of deep learning and the dynamic nature of such systems in live environments. Approaches involving collecting data from users or setting up sample profiles in sock-puppet audits can have limitations in terms of scaling to very large samples~\cite{adalovelace2021}. To overcome these limitations and provide scalable and comprehensive risk assessments this project developed a simulation environment. These environments allow researchers to systematically measure user engagement and assess the societal impact of recommender algorithms in a controlled setting. They can also address issues pertaining to user data in research through the generation of synthetic user data. Simulation frameworks such as Google's RecSim \cite{ie2019recsim}, Alibaba's Virtual-Taobao \cite{shi2019virtual} and Microsoft's MindSim \cite{luo2022mindsim} have been developed primarily for use by developers of recommendation algorithms to test and refine their systems. This project builds on this work developing a prototype simulation environment for auditors,  researchers and industry to assess risk in recommender algorithms.


%

\section{Evaluation Platform Design}

\begin{figure}[!ht]
    \centering
    \includegraphics[width=1.05\linewidth]{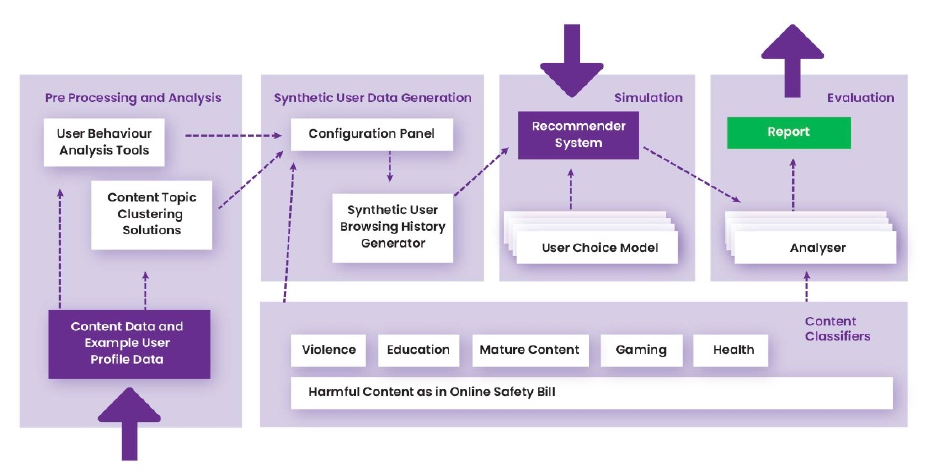}
    \caption{ARTAI Components}
    \label{fig:framework}
\end{figure}


ARTAI is designed as a simulation platform for recommendation algorithms to impart transparency and enable large-scale evaluation of trends in what content is suggested to users online. The simulation environment is integrated within a user interface designed for platform regulators, third-party auditors and industry. It allows for sample profiles of users and different models of online interaction to be generated and interaction with the recommendation algorithm simulated. The output of the algorithms, the recommended content, is categorised and trends are compiled in an interpretable report which visualises what kind of content algorithms are recommended. This approach allows for the identification of both instances of harmful content and also potentially harmful content distribution trends. The overall aim is that societal risks, particularly risks to vulnerable groups may be identified before the negative consequences become evident in society, thus reflecting a core aim of the DSA which is to ``foresee'', rather than witness societal risk. 

The platform comprises five main components: \textit{Pre-Processing and Analysis}, \textit{Content Classifiers}, \textit{Synthetic User Data Generation}, \textit{Simulation} and \textit{Risk Evaluation} (see Figure \ref{fig:framework}). 

\textbf{Pre-Processing and Analysis}: Given a large-scale dataset such as videos or text, along with profile and online behaviour data of users, this component provides a suite of tools to pre-process the data and uncover patterns and trends that can be used to generate data synthetically. For instance, user behaviour analysis tools track and analyse interactions across various dimensions such as engagement frequency and content preferences, to uncover behavioural trends. Content topic clustering solutions categorise and group content based on topics to reveal underlying structures within the data. These insights can then inform the generation of large-scale synthetic user data for simulations. In our work, we utilised a recently released large-scale dataset of short videos MicroLens \cite{ni2023content}. This dataset includes extensive user-item interaction data along with rich visual and textual information such as the original video content, comments and titles. These details enable a more fine-grained classification of the items and facilitate the learning of user preference distributions for data generation.

\textbf{Content Classifiers}: This component includes a number of trained classifiers designed to identify various types of videos or text. The primary goal is to discern the kind of content being recommended, rather than merely identifying harmful content. The content classification tool developed by ARTAI incorporates advanced natural language processing (NLP) technologies to ensure that classification outcomes are both explainable and meaningful. This approach enhances the transparency and utility of the tool, enabling users to understand and trust the classification process. It also allows researchers studying the social impact of recommender algorithms to define and investigate new topics.

\textbf{Synthetic User Data Generation}: This component involves example user groups that can be specified and synthetic data generated. Groups of users can be set up with different interest distributions. For instance, we may be interested in users who have the same interests but marginal differences in their browsing histories in relation to a particular topic. Setting up this interest distribution will allow for the simulation of recommender system outputs to capture the consequences of even marginal changes in online behaviour for user groups. 

\textbf{Simulation}: The use of simulation is crucial for understanding how different recommendation strategies influence user behaviour and content dissemination over time. By creating a controlled environment, we can systematically analyse the impact of various factors and conditions on the outcomes of recommender systems. Multiple user choice models are represented in the platform \cite{hazrati2022simulating} and this is used to simulate the interaction between users and recommendation algorithms. For example, in a position-based user choice model, users are most likely to select items recommended at the top of a personalised recommendation list. How actively a user interacts with recommender systems over time will also be included as part of the modelling.

\textbf{Risk Evaluation}: The platform will develop reports based on collated patterns and trends of recommended content for synthetic user data, simulated interactions and the kind of personalised recommendations suggested to users over time. This component will highlight how certain user profiles and online behaviours give rise to different content recommendations. The content classifiers allow for the evaluation, not only of what conditions may lead to the recommendation of particular harmful content types, but also of the proportion of content categories that are recommended overall and how this changes over time.

\section{Conclusion}

This paper presented ARTAI, a simulation environment for the evaluation of recommender algorithms for societal risk. The platform aims to address the lack of oversight into what kind of content is being recommended to different user groups, which is a particular issue concerning vulnerable groups such as children. ARTAI is designed primarily for use by auditors and vetted researchers to enable the implementation of the European Union DSA. While simulation environments have been used in industry (e.g.  \cite{ie2019recsim,shi2019virtual,luo2022mindsim}), they have been created with developers in mind and require advanced technical skill to implement and use. The ARTAI system is designed from the outset with the full range of stakeholders involved resulting in an accessible user interface. The goal of this project is to equip auditors with tools to conduct ethics evaluations and increase the scale at which researchers across disciplines can evaluate the social impact of online platforms. 


\begin{acks}
This publication has emanated from research conducted with the financial support of the EU Commission Recovery and Resilience Facility under the Science Foundation Ireland OurTech Challenge Grant Number 22/NCF/OT/11077 and with the financial support of Science Foundation Ireland 12/RC/2289\_P2 at Insight the SFI Research Centre for Data Analytics at University College Dublin.
\end{acks}

\bibliographystyle{ACM-Reference-Format}
\input{Fac24.bbl}


\end{document}